# Image-Based Alzheimer's Disease Detection Using Pretrained Convolutional Neural Network Models

Nasser Alsadhan

*Department of Computer and Information Sciences, King Saud University, Riyadh, Saudi Arabia*

**Abstract:** Alzheimer's disease is an untreatable, progressive brain disorder that slowly robs people of their memory, thinking abilities, and ultimately their capacity to complete even the most basic tasks. Among older adults, it is the most frequent cause of dementia. Although there is presently no treatment for Alzheimer's disease, scientific trials are ongoing to discover drugs to combat the condition. Treatments to slow the signs of dementia are also available. Many researchers throughout the world became interested in developing computer-aided diagnosis systems to aid in the early identification of this deadly disease and assure an accurate diagnosis. In particular, image-based approaches have been coupled with machine learning techniques to address the challenges of Alzheimer's disease detection. This study proposes a computer-aided diagnosis system to detect Alzheimer's disease from biomarkers captured using neuroimaging techniques. The proposed approach relies on deep learning techniques to extract the relevant visual features from the image collection to accurately predict the Alzheimer's class value. In the experiments, standard datasets and pre-trained deep learning models were investigated. Moreover, standard performance measures were used to assess the models' performances. The obtained results proved that VGG16-based models outperform the state-of-the-art performance.

## Introduction

Alzheimer's Disease (AD) is a form of dementia that causes memory, cognitive and behavioral impairments. Symptoms often start gradually and intensify over time, eventually becoming severe enough to disrupt everyday activities (Alzheimer's Association, 2019).

Alzheimer's disease is not an inevitable part of aging. However, one of the primary causes of its manifestation is advancing age. Notably, the majority of Alzheimer's patients are 65 and older. Furthermore, Alzheimer's disease does not appear suddenly; dementia symptoms emerge gradually. Memory loss is limited in the beginning stages of Alzheimer's, but people with late-stage Alzheimer's frequently lose their capacity to communicate and respond to their surroundings (Alzheimer's Association, 2019). Even with advances in neuroimaging techniques, physicians and doctors have difficulty diagnosing Alzheimer's disease stages. Approximately 5.8 million Americans of all ages lived with Alzheimer's dementia in 2019. The distribution of this is given in Fig. 1. Manual diagnosis of Alzheimer's disease is subjective and time-consuming and geriatricians are sometimes needed to determine the exact stage of the disease. Computer Aided Diagnosis (CAD) can help physicians and geriatricians make the right decisions and determine the patient's exact stage. Some works have made use of deep learning models (Khojaste-Sarakhsi *et al.*, 2022). Deep Learning methods can assist in identifying the best features for describing the patient's condition, resulting in a more rapid and effective diagnosis, saving countries trillions of dollars in medical care costs. In this study, the researchers design a novel CAD system using the latest advancement in deep learning to detect the current Alzheimer's stage of a patient. In addition, the performance of the proposed system is assessed using standard metrics, datasets, and state-of-the-art CAD systems.

The issue with conventional CAD systems is that feature extraction must be carried out manually and is time-consuming. In this study, we proposed a method for accurately predicting the Alzheimer's class value from an image collection by extracting the pertinent visual features using deep learning techniques.

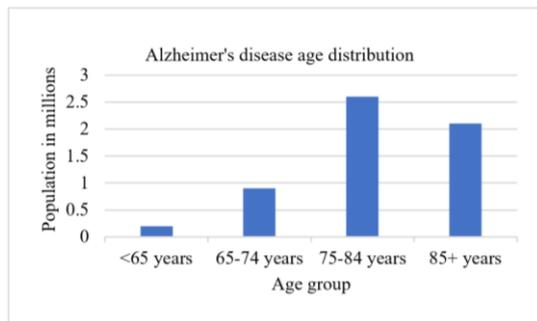

**Fig. 1:** Age distribution of people with Alzheimer's dementia in the US

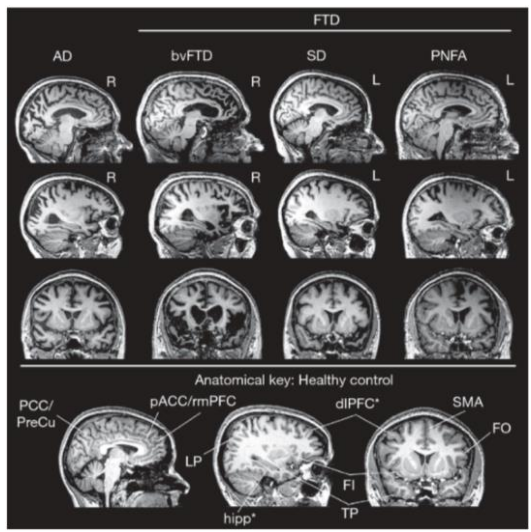

**Fig. 2:** Sample MRI of patients with AD

## Background

### Types of Alzheimer's Disease

There are three types or severities of Alzheimer's disease, which we outline (Husebo *et al.*, 2016).

### Mild Alzheimer's

At this stage, patients encounter difficulties doing daily chores such as work tasks and bill payments, among other things. These symptoms are not very serious, but at this stage, patients can work with some difficulty; they take longer than usual to complete everyday tasks, which they used to do with ease previously.

### Moderate Alzheimer's

Patients in this stage are more vulnerable to neuronal damage, resulting in more severe Alzheimer's disease symptoms. Because of the memory loss, the confusion worsens and they require assistance from others. These individuals, despite their physical agility, are unable to accomplish everyday activities because their delusions take over the sensory processing of their thinking.

### Severe Alzheimer's

The brain cells start dying as the plaques and tangles spread, reducing the brain tissue's size. Patients with this condition usually stay in bed and cannot communicate normally.

### Symptom Stages of Alzheimer's Disease

Generally, the symptoms of Alzheimer's disease are divided into three main stages (Kim *et al.*, 2021).

### Early Symptoms

Memory loss is the primary symptom of Alzheimer's disease in its early stages. For instance, forgetting the names of places and items, having trouble finding the correct term, and continuously asking inquiries. Increased anxiety, moments of bewilderment, and behavioral abnormalities are also prevalent.

### Middle-Stage Symptoms

Memory issues will worsen as Alzheimer's disease progresses. Some patients in this stage may struggle to remember the names of people they know and not recognize family and friends. At this stage, some Alzheimer's patients typically require assistance with daily living activities from others. They may require assistance with eating, washing, dressing, and using the toilet.

### Later Symptoms

Alzheimer's symptoms will worsen as the disease progresses, which can be worrisome for patients, carers, friends, and family. Delusions and paranoia may occur and fade during the disease, but they might intensify as it advances. People with advanced Alzheimer's disease may require full-time care and assistance with eating, moving, and personal care.

### Neuroimaging

Neuroimaging is a clinical specialty that focuses on creating images of the brain using noninvasive techniques (e.g., magnetic resonance imaging and computed tomography). Brain imaging has become possible since the discovery of X-rays and with new technologies and large-scale imaging projects such as the Human brain project, Human connectome project, and Chinese brain project, more detailed information about the brain has been obtained (Bilgiç, 2018).

Magnetic Resonance Imaging (MRI) is preferable to Computerized Tomography (CT) generally because it is more sensitive to minuscule changes in tissue signal and is superior to CT in the initial and follow-up examination of enhancing lesions such as neoplasms, infection, and vascular lesions (Britton, 2016). Figure 2 gives an example of an MRI.

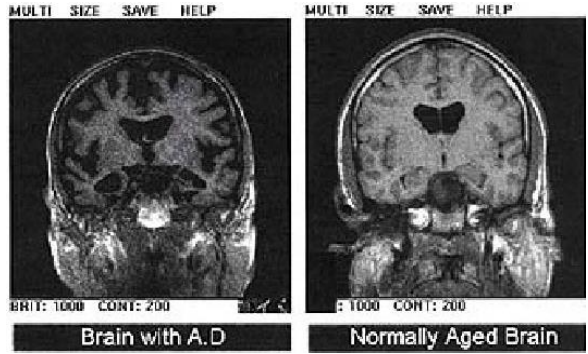

**Fig. 3:** A comparison between MRIs of a brain with AD and a normal brain

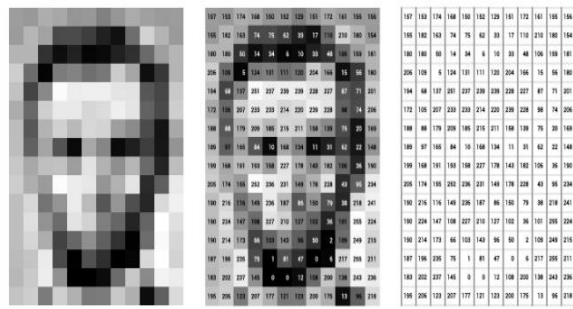

**Fig. 4:** Greyscale image pixel values are stored in a matrix. Each element of the matrix determines the intensity of the corresponding pixel, using a number between 0 (indicating black) and 255 (indicating white)

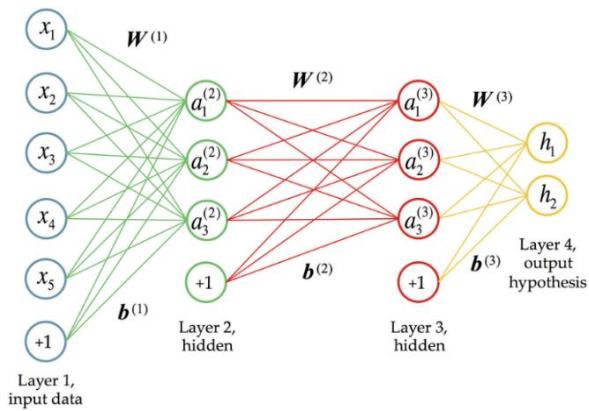

**Fig. 5:** A deep neural network with two hidden layers is illustrated. The input data is represented by the first layer (each dataset comprises five statistics), and the last layer predicts the two response variables. The bias term is represented by the final node (+1) in each input layer

### MRI

Magnetic Resonance Imaging (MRI), a medical imaging technology, employs a field of magnets and radio waves generated by a computer to obtain exact pictures of the tissues and organs in your body. MRI scans create a detailed image of internal organs and structures using a large magnet, radio waves, and a computer. The scanner resembles a tube with a table in the center onto which the patient can slide. MRI scans are distinct from CT and X-rays because they do not use potentially harmful ionizing radiation. Mass assessment in various planes is possible with MRI, which is very valuable in surgical or stereotactic irradiation planning. In this study, we will train our deep-learning model with MRI scans (Britton, 2016; Matsumoto *et al*., 2015). Figure 3 shows the difference in MRI between a person with AD and a healthy person.

### Greyscale Imaging

Grayscale is a collection or range of monochromatic (gray) colors ranging from absolute white on one end of the spectrum to absolute black on the other. Color and brightness information, or greyscale, are typically included in images. A digital image comprises groups of Red, Green, and Blue colors (RGB), so when color is removed, the image becomes greyscale. Figure 4 gives an example of a greyscale image.

### Supervised Learning

Supervised learning, often known as supervised machine learning, is a machine learning and artificial intelligence subfield. It differentiates itself by training algorithms on labeled datasets to accurately identify data or forecast outcomes.

### Classification

The Classification algorithms will take all input data and assign it to the appropriate category or class based on the labeled data provided at the start. For example, if a picture of an animal is provided as input, the classification algorithm will determine whether it is for a dog, a cat, or a bird. The features from the input will be fed into the model, which will search for relationships between the input and each class in the data set.

The classification mapping function is $Y = f(x)$, where $Y$ is the predicted output determined by the mapping function that assigns a class to an input value $x$.

### Deep Learning

Deep learning is an area of machine learning and one of the components of artificial intelligence. It replicates the human brain in data processing particularly when it involves decision-making.

### Neural Network

Neural networks are a collection of algorithms that identify patterns and are based on the human brain. It can predict patterns based on training input data such as images, sounds, and texts. The input layer is usually the first layer in a neural network structure, followed by hidden layers and finally, an output layer to classify the input (Abiodun *et al*., 2018). Fig. 5 gives an example of a neural network.

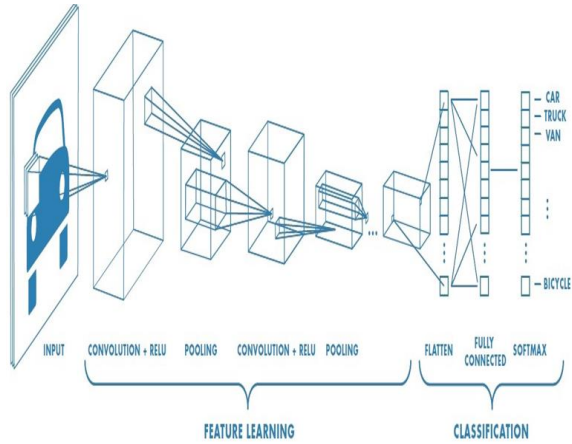

**Fig. 6:** Neural network with many convolutional layers

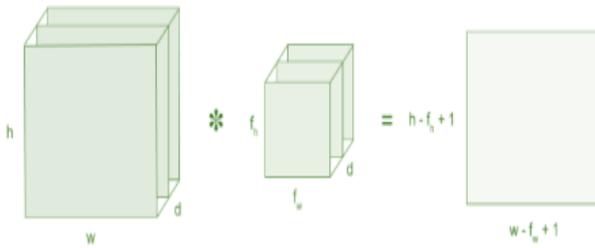

**Fig. 7:** Image matrix multiplies kernel or filter matrix

| Operation | Filter | Convolved Image |
|---|---|---|
| Identity | $\begin{bmatrix} 0 & 0 & 0 \\ 0 & 1 & 0 \\ 0 & 0 & 0 \end{bmatrix}$ | |
| Edge detection | $\begin{bmatrix} 1 & 0 & -1 \\ 0 & 0 & 0 \\ -1 & 0 & 1 \end{bmatrix}$ | |
| | $\begin{bmatrix} 0 & 1 & 0 \\ 1 & -4 & 1 \\ 0 & 1 & 0 \end{bmatrix}$ | |
| | $\begin{bmatrix} -1 & -1 & -1 \\ -1 & 8 & -1 \\ -1 & -1 & -1 \end{bmatrix}$ | |
| Sharpen | $\begin{bmatrix} 0 & -1 & 0 \\ -1 & 5 & -1 \\ 0 & -1 & 0 \end{bmatrix}$ | |
| Box blur (normalized) | $\frac{1}{9}\begin{bmatrix} 1 & 1 & 1 \\ 1 & 1 & 1 \\ 1 & 1 & 1 \end{bmatrix}$ | |
| Gaussian blur (approximation) | $\frac{1}{16}\begin{bmatrix} 1 & 2 & 1 \\ 2 & 4 & 2 \\ 1 & 2 & 1 \end{bmatrix}$ | |

**Fig. 8:** Standard filters of CNN

## Convolutional Neural Network

Convolutional Neural Networks (CNNs) are most typically used for image recognition, image classification, object identification, and face recognition. CNNs take an image as input and process it before categorizing it (dog, cat, cow). CNNs take images as input, which are arrays of pixels whose size is determined by their resolution. The primary application of CNN is to train convolutional layers with filters, pooling, and fully connected layers to identify image properties (Gu *et al*., 2018). Figure 6 depicts the CNN model's process.

## Convolution Layer

The Convolution Layer is the first layer that extracts features from the input image. After extracting image features, it maintains pixel relationships and then learns image features using small squares of input data. It is a logical operation with two inputs: An image matrix and a filter or kernel. A volume dimension (height-fh +1) * (weight-fw +1) *1 is produced by multiplying an input image matrix (volume) of dimension (height * weight * dimension) by an input filter (fh * fw * d). This is illustrated in Fig. 7. Images can be processed using various filters, which can blur, sharpen, or detect image edges. Figure 8 depicts a convolution image that has been processed using various filters.

## Pooling Layer

When the input picture is too huge, the Pooling layer's function is to minimize the number of parameters. Spatial pooling, also known as subsampling or down sampling, decreases the dimensionality of each map while keeping important information (Gu *et al*., 2018). There are several types of spatial pooling:

- Max pooling takes the most significant element from the rectified feature map created after using a ReLu
- Average pooling uses the rectified feature map's average
- Sum pooling: Sums all elements in the rectified feature map

## Non-Linearity (ReLU)

The activation function ReLU, which stands for "Rectified Linear Unit," is written as:

$$f(x) = \max(0, x) \qquad (1)$$

The importance of ReLU is to keep the convolutional neural network from learning negative linear values. ReLU is one of several activation functions that can be used, such as tanh or sigmoid. However, the performance of ReLU is better (O'Shea and Nash, 2015). Figure 9 illustrates the ReLU operation.

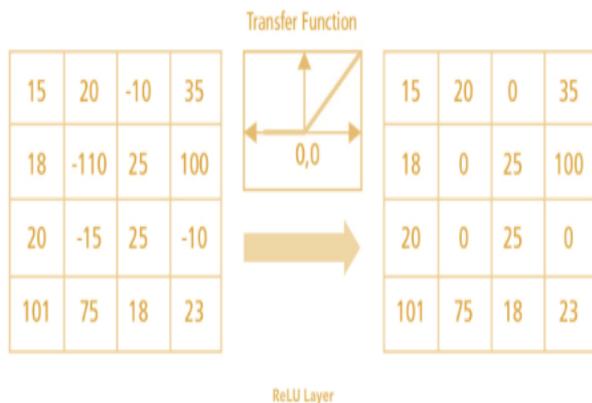

**Fig. 9:** ReLU operation

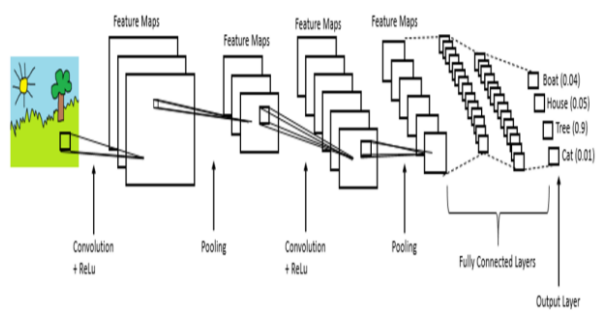

**Fig. 10:** A complete CNN architecture

*Fully Connected Layer*

In the fully connected layer, the matrices are converted to vectors and loaded into a fully connected layer, similar to a neural network. Following these processes, we combine the features to form a model. In this model, an activation function like softmax or sigmoid is used to determine whether the output image is of a dog, a cat, or a car. Figure 10 gives the complete architecture of a CNN.

Some research papers have used artificial intelligence to solve the AD detection problem. Shallow model-based approaches and deep learning-based models were the most commonly used approaches (Feng *et al*., 2019).

## Literature Review

*Deep Learning-Based Solutions*

A particular type of machine learning called deep learning has the potential to resolve MRI-based classification issues like Alzheimer's disease detection. Deep learning MRI scores surpassed virtually all other biomarkers in diagnosing prodromal illness and predicting clinical development, including, surprise, indicators of amyloid or tau pathology (Basaia *et al*., 2019). The damage caused by Alzheimer's disease is visible using a variety of imaging modalities, including PET and two types of MRI, structural and functional Magnetic Resonance Imaging (sMRI, fMRI), and other methods of imaging the brain. Huang *et al*. (2019) used a Convolutional Neural Network (CNN) to combine multimodal information from T1-MR and FDG-PET images of the hippocampal area to detect AD disease. They also used 3D image-processing CNNs, so they do not require extracting features manually.

Similarly, several studies have used deep learning to detect Alzheimer's disease (Basaia *et al*., 2019). Deep learning has the advantage over shallow models because it can model and categorize patterns to extract novel features. Some authors have linked sparsity penalties to the standard neural network training method (Jie *et al*., 2016). A convex relaxation method was introduced by Feng *et al*. (2019) to solve optimization problems involving sparsity. However, in terms of time complexity, this resulted in a costly algorithm. Teikari *et al*. (2016) proposed complex non-linear activation brain modeling. To improve classification performance, parametric non-linear encoding functions were used. Using Alzheimer's Disease Neuroimaging Initiative (ADNI) MRI and FGD-PETE data sets, the researchers (Mehmood *et al*., 2020) presented a multimode cascaded convolutional neural network. Similarly, the authors (Jyoti and Zhang, 2018) used the ADNI MRI dataset to create a 3D convolutional neural network and used an end-to-end training method to classify Alzheimer's disease cases.

Additionally, a number of researchers used imaging methods like diffuser tensors and functional Magnetic Resonance Imaging (fMRI) to study connectivity. Previous research in causal references and tractography in Alzheimer's disease demonstrated that behavioral changes in patients are associated with changes in connectivity. Other researchers Liu *et al*. (2014b) linked deep learning to a zero-masking strategy to improve detection performance in multiclass Alzheimer's disease diagnosis.

Finding large-scale datasets is a problem for deep learning in all fields and some techniques for data augmentation have been developed. We employ a novel data augmentation strategy well-suited for MRI-only datasets, including scans obtained from patients over multiple visits (Matsumoto *et al*., 2015). Another critical factor in deep learning resources is the device used to train the model for the CAD system. Because neural network training is powerful, it must be done on a GPU, or it will take a prolonged time to train.

*Shallow Models-Based Solutions*

Aside from deep learning, there are numerous techniques for detecting Alzheimer's disease, such as machine learning and other frameworks. Some of them work on feature selection; for example, Ding *et al*. (2015), improved feature subset selection by combining SVM-RFE and covariance to capture the relationship between features. In contrast, the authors of (Liu *et al.*, 2014a) presented a sparse learning method with tree-structured regularization for recognizing pathological degeneration patterns at fine to coarse scales. This was done to better diagnose disease by identifying informative imaging biomarkers. In Chincarini *et al.* (2011), a multimodal classification approach based on multiple-kernel SVM was outlined. Biomarkers, including sMRI, Cerebrospinal Fluid (CSF), and Positron Emission Tomography (PET), were used to detect AD cases. This two-class classification approach yielded promising results. An SVM-based classification strategy for multiclass classification was recently described by Beheshti *et al.* (2015). In this study, just a subset of characteristics from structural MRI was retrieved and put into a kernel logistic regression. This was done to lower the computational cost of the suggested method.

In addition, Laila *et al.* (2015), the researchers proposed using different kernel-based multiclass SVM classifiers to differentiate between the two AD classes. Furthermore, they used a 10-fold stratified cross-validation strategy to help determine the classification model's optimal hyperparameters. Their experiments demonstrated that combining multimodal biomarkers outperforms approaches that use a single data modality of biomarkers.

Reviewing several research papers revealed that techniques other than deep learning can be complicated because they require manual work such as feature extraction, feature selection, and more. Deep learning, on the other hand, is much simpler. Table 1 summarizes the condition of the work described above:

**Table 1:** Different classification techniques belonging to studies that utilize MR images

| Ref | Preprocessing | Feature selection | Classifier | Accuracy |
|---|---|---|---|---|
| Feng *et al.* (2019) | Spatial normalization | - | CNN | 0.970 |
| Islam and Zhang (2018) | Batch normalization | - | CNN SoftMax layer | - |
| Oh *et al.* (2019) | - | - | CNN | 0.860 |
| Basaia *et al.* (2019) | - | - | CNN | 0.980 |
| Oh *et al.* (2019) | - | - | SVM | 0.790 |
| Mehmood *et al.* (2020) | - | - | VGG-16 | 0.990 |
| Jyoti and Zhang (2018) | Gray-matter segmentation | - | SAE | 0.820 |
| Chincarini *et al.* (2011) | Segmentation | - | multi-kernel multimodal classification | 0.900 |
| Beheshti *et al.* (2015) | Spatial normalization+ Intensity normalization | Relevance measured by random forest | SVM-20fold CV | - |
| Beheshti *et al.* (2016) | VBM | Probability distribution Function | SVM (linear)-10 fold CV | 0.940 |
| Liu *et al.* (2014a) | VBM | t-test feature ranking | SVM (linear)-10fold CV + fisher criterion | 0.896 |
| Laila *et al.* (2015) | Intensity normalization + Tissue segmentation + Spatial normalization | tree-guided sparse learning | SVM (linear) -10fold CV | 0.902 |
| Casanova *et al.* (2013) | Spatial normalization + Tissue segmentation | Partial least square | SVM (linear) -10fold CV | 0.884 |
| Ding *et al.* (2015) | Tissue segmentation + Intensity normalization | - | Regularized logistic regression 10-fold CV | 0.871 |
| Cho *et al.* (2012) | VBM | Recursive feature Elimination + covariance | SVM (RBF) | 0.928 |
| Dubey *et al.* (2014) | Spatial normalization | Principal Component Analysis (PCA) | Linear Discriminant Analysis (LDA) | - |
| Herrera *et al.* (2013) | Cortical reconstruction + Tissue segmentations | Sparse logistic regression | Random forest | 0.872 |
| Ahmed *et al.* (2015) | Spatial normalization | - | SVM (RBF) -10fold CV | 0.962 |
| Liu *et al.* (2018) | Spatial normalization + Segmentation | PCA | SVM (RBF) -10fold CV | 0.837 |
| Wang *et al.* (2018) | - | - | SoftMax layer | 0.933 |
| de Vos *et al.* (2018) | Motion correction | LDA | AdaBoost ensemble | 0.758 |
| Kazemifar *et al.* (2017) | Brain extraction + Intensity normalization + Spatial normalization | - | Elastic net logistic regression | 0.780 |
| Petrick *et al.* (2013) | Spatial normalization + Smoothing | Goodness-of-Fit (GoF) calculation + template matching + SVM classification | SVM (linear) Leave-one-out CV | 0.750 |

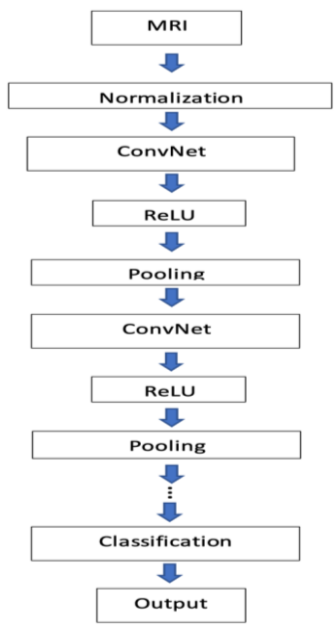

**Fig. 11:** Proposed CNN model

## Materials and Methods

### Proposed Method

This section describes (i) The proposed system and (ii) The CNN architecture used to solve the classification model. An overview of the proposed CAD systems will be provided, followed by a depiction of the proposed CNN model.

### System Overview

Computer-Aided Diagnosis (CAD) assists doctors in interpreting medical images such as X-rays and MRIs. More than ever, CAD systems assist physicians and doctors in detecting illnesses and diseases. Computer-aided detection systems detect portions of an image that may indicate certain problems and notify doctors of them throughout image interpretation. In current practice, as mandated by the FDA, a physician must first interpret the image before the CAD system can be used (Castellino, 2005). A CAD algorithm that uses machine learning requires an image dataset to analyze, in our case, an MRI dataset.

### Proposed CNN-Based Model

Initially, the motivation for using Convolutional Neural Network (CNN) was that it is primarily used for image recognition, image classifications, object detection, and face recognition. The proposed model is depicted in Fig. 11.

The MRI layer represents the labeled data that the model will learn. The normalization layer's purpose is to keep the learning process stable. The feature extraction takes place in the Convolution layer. Furthermore, the ReLU is a transfer function activation function, which prevents the model from learning negative values. The Pooling layer handles image size while retaining the essential features. The fully connected layer in the classification layer converts the feature map matrix to a vector before creating the model. Finally, the softmax activation function is used to categorize the output.

### Transfer Learning

A large-scale labeled database is required to train CNNs. This is a significant challenge for our application because it currently requires a public database with many labeled images. To address this limitation, we propose transfer learning, which entails pre-training a model on an unrelated task with an extensive existing database and reusing that model in our paper. Several pre-trained CNN models have been released. LeNet, AlexNet, VGGNet, GoogLeNet, and ResNet are all widely used for various applications. Some are obsolete, while others have evolved. We will investigate the best VGGNet and ResNet in this study, as most papers in this field use these two architectures.

### Experimental Design

We started with a public dataset from Kaggle (Dubey, 2020), which comprises two files. A test dataset of 1279 MRIs and a training dataset of 5121 MRIs, for a total of 6400 MRIs, each containing four files for Alzheimer's stages. Mild Demented has 717 MRIs for training, Very Mild Demented has 1792, ModerateDemented has 52 and Non Demented has 2560. MildDemented has 179 MRIs for testing, Very Mild Demented has 448, ModerateDemented has 12 and NonDemented has 640. The information is gathered by hand from various websites and each label is verified. The NIFTI format is required by a significant number of MRI analysis tools. Our obtained dataset had previously been cleansed when we converted it to NII and subsequently to JPG format. In our case, the noise-reduced dataset needed little pre-processing, to begin with. The data from the train, validation, and test runs had been rescaled to 1./255. We then set a desired input size of 244*244 for all of the data.

### Training

We split the training dataset into two parts: Training and validation. 20% of the training dataset will be used to create the validation dataset. The validation dataset will objectively assess the proposed model's fit on the training dataset. The validation set will determine whether the model is overfitting or underfitting.

### Performance Measures

We utilize the accuracy, precision, sensitivity, and f-measure to evaluate the performance of the proposed model. Instances in a predicted class represent each column of the confusion matrix, while instances of the actual represent each row. This is shown in Table 2.

**Table 2:** Confusion matrix

|  | Predicted Positive (P) | Predicated Negative (N) |
|---|---|---|
| Actual Positive (P) | True Positive (TP) | False Negative (FN) |
| Actual Negative (N) | False Positive (FP) | True Negative (TN) |

The definitions of the terms in Table 2 are as follows:

− Positive (P) for a disease means the patient has dementia
− A Negative (N) for a disease means the patient is healthy
− If the outcome is True Positive (TP), meaning the patient has dementia and is predicted to have it
− If the outcome is True Negative (TN), the patient does not have dementia and is predicted not to have it
− If the outcome is False Positive (FP), the patient does not have dementia but is predicted to have it
− If the outcome is False Negative (FN), the patient has dementia but is predicted not to have it

The number of true predictions over the total number of predictions calculates the accuracy:

$$Accuracy = \frac{TP + TN}{TP + TN + FP + FN} \quad (2)$$

The total number of true positive predictions/total number of correct predictions calculates sensitivity:

$$Sensitivity = TPR(True\,Positive\,Rate) = Recall = \frac{TP}{TP + FN} \quad (3)$$

The number of true negative predictions/number of true negative and false positive predictions calculates specificity:

$$Specificity = \frac{TN}{TN + FP} \quad (4)$$

The total number of true positive predictions/total number of positive predictions calculates precision:

$$Precision = \frac{TP}{TP + FP} \quad (5)$$

F-measure will be closer to the smaller value of Precision or Sensitivity:

$$F - measure = 2 \times \frac{Precision \times Recall}{Precision + Recall} \quad (6)$$

*Experiments Scenarios*

We chose a dataset with four classes: Very Mild Demented (1882 MRIs), Mild Demented (754 MRIs), Moderate Demented (64 MRIs), and Non-Demented (2688 MRIs). We investigated the ResNet50 and VGG16 models, first training them on four classes with no data augmentation. ResNet50 performed poorly, but VGG16 performed noticeably better.

Furthermore, we dropped the Moderate Demented class because it only had 64 MRIs and began training the model on three classes that produced roughly the same results. We also used random cropping, horizontal and vertical flipping, grey scaling, and random rotating MRIs to supplement the dataset. Unfortunately, the results did not improve significantly. We also tried setting the number of validation MRIs at 200 for each class, whereas previously, it took 20% of the training set for validation, causing the number of MRIs in the validation sets to differ; this did not improve the results. We manually divided the training dataset into training and validation sets as a last resort.

Finally, an attempt to merge the original four classes into two classes, Demented and Non-Demented, employing all of the above modifications, including augmentation and manual splitting, yielded the best results thus far.

To determine the final parameters for the suggested model, hyperparameter tuning was done. The Adam optimizer showed promise in this tuning and it was ultimately selected to create our suggested model. Furthermore, the suggested model's loss in identifying Alzheimer's MRI pictures based on anticipated values was tested using sparse categorical cross-entropy. 70 epochs of the proposed model were run in a GPU environment on a Google Colab notebook. The model was also run with various iteration counts, but 70 iterations produced the best results. Each epoch's execution took an average of about 4 s. 128 was chosen as the batch size and this number has produced better results than the other options.

**Results**

The results of using two, three, and four classes are shown in Fig. 12. The three and four class results performed worse, owing to a lack of MRIs, whereas the two classes had far more MRIs to work with for each class. A graph showing the accuracy results across different epochs using 2 classes is given in Fig. 13.

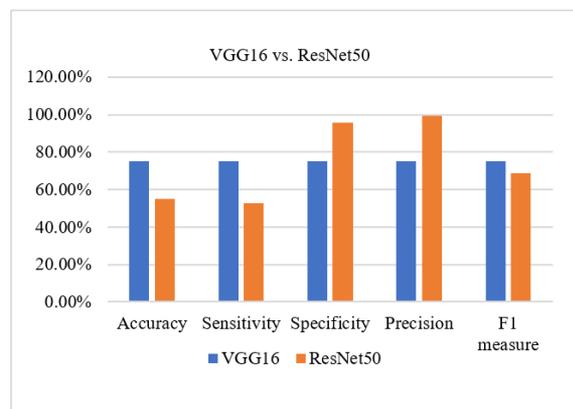

**Fig. 12:** VGG16 and ResNet50's performances in accuracy, sensitivity, specificity, precision, and F-measure in 2 class tests

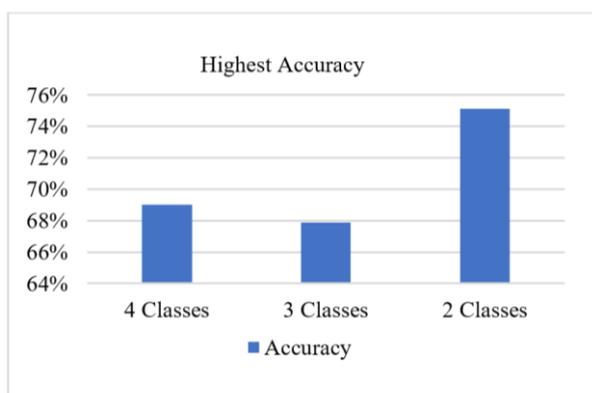

**Fig. 13:** Accuracy results using two, three, and four classes

## Discussion

As previously stated, we began training both ResNet50 and VGG16 using four classes. They both delivered results that were below expectations. ResNet50 had a 60% accuracy rate, averaging 57% across 23 tests for various hyper-parameters. The VGG16-based model, on the other hand, produced better results, achieving 69% and averaging 64% across 23 tests. The results were roughly identical after dropping the Moderate Demented class and training again. We also attempted to augment the dataset but were unsuccessful. The best results were obtained by using two classes (demented/not demented). Figure 14 shows the results. VGG16 is more consistent than ResNet50 in its performance, as it has similar scores in all performance metrics. ResNet50 on the other hand has high precision and specificity, which suggests that it is more confident in predicting the demented class but at the cost of a low recall, which in real life will affect the diagnosis of many people with dementia. For this reason, the performance of VGG16 is considered better.

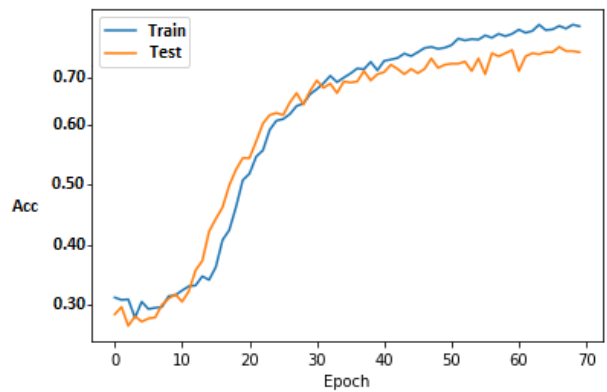

**Fig. 14:** Accuracy across different epochs using two classes

## Conclusion

Alzheimer's disease detection is a vibrant, emerging research field that faces numerous challenges. Several approaches to addressing Alzheimer's disease detection issues were investigated in this study. We developed and tested a CAD system to detect Alzheimer's disease using biomarkers captured using neuroimaging techniques. Deep learning techniques were used in the proposed approach to extract relevant visual features from the image collection and accurately predict the Alzheimer's class value. Standard datasets and pre-trained deep learning models were investigated in the experiments.

Furthermore, after determining the best model, we evaluated its performance using standard performance measures such as sensitivity, specificity, precision, and F-measure. The obtained results demonstrated that VGG16 outperforms ResNet50 and produces better results.

Finally, deep learning methods can achieve better results in this field with more MRIs, an extensive dataset, and sufficient computing power to conduct the training.

The results found in this study are promising, but one limitation is that the proposed model performs well at distinguishing two classes.

Although this study mainly concentrated on comparing two pretrained models, future work includes more extensive research on other deep learning models such as YOLO and RetinaNet in hopes of obtaining better results while also using all four classes.

## Acknowledgment

This study was supported by the Research Center of the College of Computer and Information Sciences at King Saud University, Riyadh, Saudi Arabia. The author is grateful for this support.

## Funding Information

The authors have not received any financial support or funding to report.

## Ethics

This article is original and contains unpublished material. The corresponding author confirms that all of the other authors have read and approved the manuscript and that no ethical issues are involved.